\documentclass[12pt]{iopart}

\usepackage{epsf}
\usepackage{graphicx}
\newcommand{\et}{{\it et al.~}}
\newcommand{\lie}{\mathcal{L}}

\begin{document}

\title[Binary black holes on a budget]{Binary black holes on a budget: 
Simulations using workstations}

\author{Pedro Marronetti, Wolfgang Tichy}
\address{Department of Physics,
	Florida Atlantic University, 
	Boca Raton, FL 33431, USA}

\author{Bernd Br\"ugmann, Jose Gonz\'alez, Mark Hannam, Sascha Husa, 
Ulrich Sperhake}
\address{Theoretical Physics Institute,
	University of Jena,
	07743 Jena, Germany}

\begin{abstract}
Binary black hole simulations have traditionally been computationally very 
expensive: current simulations are performed in supercomputers involving 
dozens if not hundreds of processors, thus systematic studies of the 
parameter space of binary black hole encounters still seem prohibitive 
with current technology. Here we show how the multi-layered refinement 
level code BAM can be used on dual processor workstations to simulate
certain binary black hole systems. BAM, based on the moving punctures method, 
provides grid structures composed of boxes of increasing resolution near 
the center of the grid. In the case of binaries, the highest resolution boxes 
are placed around each black hole and they track them in their orbits until 
the final merger when a single set of levels surrounds the black hole remnant. 
This is particularly useful when simulating spinning black holes since
the gravitational fields gradients are larger. We present simulations of 
binaries with equal mass black holes with spins parallel to the binary axis 
and intrinsic magnitude of $S/m^2= 0.75$. Our results compare favorably to 
those of previous simulations of this particular system. We show that the
moving punctures method produces stable simulations at maximum spatial 
resolutions up to $M/160$ and for durations of up to the equivalent of 
20 orbital periods. 
\end{abstract}


\ead{pmarrone@fau.edu}

\submitto{\CQG}

\maketitle


\section{Introduction}
\label{intro}

As the latest generation of gravitational wave detectors becomes
operational, the problem of faithfully simulating the evolution of binary
systems of compact objects, black holes in particular, has become
increasingly important. While post-Newtonian (PN) approximations can be used in
the first stages of the life of a binary black hole (BBH), when the two
objects get close and are rapidly orbiting around each other only solutions
to the full non-linear Einstein equations can provide the desired level of
precision. Due to the complexity of such equations, these solutions can only
be achieved by means of numerical algorithms. These results are of
particular interest for laser-interferometric observatories since BBH will
be highly relativistic when entering the sensitivity range of the detectors.

Such simulations pose a hard and challenging problem. Until recently they
tended to fail after a very short time due to 
instabilities \cite{Jansen:2003uh} which resulted
in exponentially growing run-away solutions. Fortunately, tremendous
progress has been achieved over the last two years \cite{Bruegmann:2003aw,
Pretorius:2005gq, Campanelli:2005dd, Baker:2005vv, Campanelli:2006gf,
Baker:2006yw, Herrmann:2006ks, Campanelli:2006uy, Sperhake:2006cy, 
Campanelli:2006fg, Baker:2006vn, Pretorius:2006tp, Bruegmann:2006at, 
Gonzalez:2006md,Tichy:2007hk}. 
Within the moving puncture approach and also with the 
generalized harmonic system, it is now possible to evolve BBH systems through 
several orbits and the subsequent merger and ringdown phases.

Stable and accurate, modern BBH simulations require large
computer resources and even modest size runs are performed on supercomputers
involving dozens or even hundreds of processors. The goal of this paper is
to showcase the ability of the code BAM \cite{Bruegmann:2006at} to evolve
certain BBH systems on workstations, providing results of comparable quality to
those obtained in simulations using much larger computer systems. Workstations
with similar characteristics to the ones used here are reasonably affordable 
(less than \$3000 USD at the time of publication). BAM
provides grid structures composed of boxes of increasing resolution near the
center of the grid. In the case of binaries, the highest resolution boxes
are placed around each black hole. The boxes track the holes in their orbits
until the final merger, when a single set of levels surround the black hole
remnant. A direct consequence of this grid structure is the efficient use of
computational resources as it will be detailed in section \ref{Perf}. BAM
currently handles fourth order accurate evolutions.

The end result of a BBH merger is a larger black hole. This final black hole
could in principle be non-spinning, however the conditions for this to occur
are very unlikely: any astrophysically realistic scenario would lead to a
spinning object. We test BAM by simulating a BBH with identical black holes
with intrinsic spins $S/m=0.75$ parallel to the orbital angular momentum.
We choose high-spin binaries since they require very high resolution near
the black holes which is currently difficult to achieve for most
numerical codes of this type. The time evolution of the BBH system is
achieved through the moving punctures method \cite{Campanelli:2005dd,
Baker:2005vv}.

In order to test the quality and accuracy of the simulations, we concentrate
on the measurement of the final black hole mass and angular momentum. To do
that, we implemented algorithms based on the conversion of surface integrals
(at the core of the definition of the ADM mass and angular momentum) to
volume integrals using Gauss' theorem. These calculations are studied and
compared with alternative ways of measuring these global quantities.

Section \ref{NM} presents a brief description of the equations and the
initial data sets and describes the details of BAM's numerical grid
structure \cite{Bruegmann:2006at} and the algorithms used to calculate the 
mass and angular momentum. Sections \ref{Perf} and \ref{Tests} presents 
performance and convergence tests respectively. Section \ref{Comp} compares 
alternative calculations of the mass and angular momentum and our results 
are discussed in section \ref{Res}.


\section{Evolution using the moving punctures method}
\label{NM}

\subsection{Initial data}
\label{sec:ID}

In order to start our simulations we need initial data for spinning
BBH with equal masses and spins.
Since we will employ the moving punctures approach
in our evolutions we will use standard puncture 
initial data~\cite{Brandt97b} with the momentum and spin parameters
in the extrinsic curvature given by 2PN estimates~\cite{Kidder:1995zr}.
It is sufficient to use 2PN estimates because standard puncture data
are inconsistent with PN theory beyond 
$(v/c)^3~\cite{Tichy02,Yunes:2006iw,Yunes:2005nn}$.
These parameters along with 2PN estimates for 
ADM mass $M^{ADM}_{\infty}$, ADM angular momentum $J^{ADM}_{\infty}$
and angular velocity $\Omega$ are shown in Table~\ref{punc_par_tab}.

\begin{table}
{\small
\begin{tabular}{c|c|c|c|c|c|c}
$m_b/M$ & $D/2M$ & $P/M$ & $S/M^2$ & $M^{ADM}_{\infty}/M$ & $J^{ADM}_{\infty}/M^2$ & $M\Omega$ \\
\hline
\makebox{\rule{0mm}{4mm}0.32555}	& 3.0000  & 0.12756 & 0.18750 & 0.98313 & 1.14034 & 0.055502\\
\hline\hline
\makebox{\rule{0mm}{4mm} $m_b/M^{ADM}_{\infty}$} & $D/2M^{ADM}_{\infty}$ & $P/M^{ADM}_{\infty}$ & $S/{M^{ADM}_{\infty}}^2$ & $M/M^{ADM}_{\infty}$ & $J^{ADM}_{\infty}/{M^{ADM}_{\infty}}^2$ & $M^{ADM}\Omega$ \\
\hline
\makebox{\rule{0mm}{4mm}0.33114} & 3.0515  & 0.12975 & 0.19399 & 1.01716 & 1.17981 & 0.054566\\
\end{tabular}
}
\caption{\label{punc_par_tab}
Initial data parameters. Here $m_b$ is the bare mass parameter of each
puncture and $M = 2 m$ is the sum of the ADM masses $m$ measured at each puncture. 
The holes have coordinate separation $D$, with puncture 
locations $(0, \pm D/2 ,0)$, linear momenta $(\mp P, 0, 0)$, 
and spins $(0, 0, S)$ with $S/m^2=0.75$. We also list the 2PN estimates for the ADM 
mass $M^{ADM}_{\infty}$, the ADM angular momentum $J^{ADM}_{\infty}$
and the angular velocity $\Omega$. These quantities are shown using two different 
scaling factors ($M$ and $M^{ADM}_\infty$) for easier comparison with work
done by other groups.
}
\end{table}

The coordinate distance $D$ and the momentum and spin parameters $P$ and $S$ 
directly enter the Bowen-York extrinsic curvature, while the bare mass
parameter is obtained from the condition that
the ADM masses measured at each puncture should be $m=M/2$.
This implies that each black hole has an individual spin
of $\frac{S}{{m}^2} = \frac{S}{(M/2)^2} = 0.75$, where
as in~\cite{Tichy03a,Tichy:2003qi,Ansorg:2004ds} we assume that $m$ 
is a good approximation for the initial individual black hole masses.
Note that these data are very close to the values
used by Campanelli \et ~\cite{Campanelli:2006uy}. If we express everything in terms of
the PN ADM mass, the largest difference is that
our bare mass parameter is about 1\% lower.

To complete the definition of the initial data, we also need to specify
initial values for the lapse $\alpha$ and shift
vector $\beta^i$. At time $t=0$ we use
\begin{eqnarray}
\alpha &=& \left(1 + \frac{m_b}{r_1} + \frac{m_b}{r_2}\right)^{-2}, \nonumber\\
\beta^i &=& 0, \nonumber
\end{eqnarray}
where $r_A$ is the coordinate distance from puncture $A$.
Both lapse and shift are updated by evolution equations depending on
the physical variables, as described below.

\subsection{Evolution of gravitational fields}
\label{EE}

We evolve the initial data with the BSSN
system~\cite{Shibata:1995,Baumgarte:1998te}. 
In the case of BSSN, the $3$-metric $g_{ij}$ is written as
\begin{eqnarray}
\label{confdec}
g_{ij} = e^{4\phi} \tilde{g}_{ij} \nonumber
\end{eqnarray}
where the conformal metric $\tilde{g}_{ij}$ has unit determinant.
In addition, the extra variable
\begin{eqnarray}
\tilde{\Gamma}^i = - \partial_j \tilde{g}^{ij} \nonumber
\end{eqnarray}
is introduced where $\tilde{g}^{ij}$ is the inverse of
the conformal metric.
Furthermore, the extrinsic curvature is split into its trace 
free part $\tilde{A}_{ij}$ and its trace $K$, and given by
\begin{eqnarray}
K_{ij} = e^{4\phi} 
         \left( \tilde{A}_{ij} + \frac{K}{3} \tilde{g}_{ij} \right) .\nonumber
\end{eqnarray}
These variables are evolved using 
\begin{eqnarray}
	\partial_0 \phi &=& - \frac{1}{6}\alpha K,\nonumber
\\
	\partial_0 \tilde g_{ij} &=& -2\alpha\tilde A_{ij}, \nonumber
\\
	\partial_0 \tilde A_{ij} &=& e^{-4\phi}
	[-D_iD_j\alpha + \alpha R_{ij}]^{TF}
\nonumber
\\
	&& + \alpha(K\tilde A_{ij} - 2 \tilde A_{ik}\tilde {A^k}_j), \nonumber
\\
	\partial_0 K &=& -D^iD_i\alpha + 
	\alpha(\tilde A_{ij}\tilde A^{ij} + \frac{1}{3} K^2), \nonumber
\\
	\partial_t \tilde\Gamma^i &=& \tilde{g}^{ij} \partial_j \partial_k
        \beta^i + \frac{1}{3} \tilde{g}^{ij} \partial_j \partial_k \beta^k +
        \beta^j \partial_j \tilde{\Gamma}^i  \nonumber \\ 
	&&  - \tilde{\Gamma}^j \partial_j \beta^i + \frac{2}{3}
        \tilde{\Gamma}^{i} \partial_j \beta^j - 2 \tilde{A}^{ij} \partial_j
        \alpha \nonumber  \\ 
	&& + 2 \alpha \left (\tilde{\Gamma}^i_{jk} \tilde{A}^{jk} + 6
          \tilde{A}^{ij} \partial_j \phi - \frac{2}{3} \tilde{g}^{ij}
          \partial_j K \right) , \nonumber
\end{eqnarray}
where $\partial_0 = \partial_t - \lie_\beta$, $D_i$ is the covariant
derivative with respect to the conformal metric $\tilde{g}_{ij}$, and ``TF''
denotes the trace-free part of the expression with respect to the 
{\it physical} metric, $X_{ij}^{TF} = X_{ij} - \frac{1}{3} g_{ij} X_k^k$.
The Ricci tensor $R_{ij}$ is given by
\begin{eqnarray} 
R_{ij} & = & \tilde{R}_{ij} + R^{\phi}_{ij} \nonumber\\
\tilde{R}_{ij} & = &   - \frac{1}{2} \tilde{g}^{lm} \partial_l \partial_m
\tilde{g}_{ij} + \tilde{g}_{k(i} \partial_{j)} \tilde{\Gamma}^k +
\tilde{\Gamma}^k \tilde{\Gamma}_{(ij)k} +        \nonumber   \nonumber\\ 
                       &  & \tilde{g}^{lm} \left( 2 \tilde{\Gamma}^k_{l(i}
                         \tilde{\Gamma}_{j)jm} + \tilde{\Gamma}^k_{im}
                         \tilde{\Gamma}_{klj} \right), \nonumber\\ 
R^{\phi}_{ij}  & = & - 2 D_i D_j \phi - 2 \tilde{g}_{ij} D^k D_k \phi + 4 D_i
\phi D_j \phi - \nonumber \\ 
                        &    & 4 \tilde{g}_{ij} D^k \phi D_k \phi. \nonumber
\end{eqnarray} 
The Lie derivatives of the tensor densities $\phi$, $\tilde{g}_{ij}$ and
$\tilde{A}_{ij}$ (with weights $1/6$, $-2/3$ and $-2/3$) are 
\begin{eqnarray*}
\lie_\beta \phi & = & \beta^k \partial_k \phi + \frac{1}{6} \partial_k \beta^k, 
\nonumber\\
\lie_\beta \tilde{g}_{ij} & = & \tilde{g}_{ij} \partial_k \tilde{g}_{ij} +
\tilde{g}_{ik} \partial_j \beta^k + \tilde{g}_{jk} \partial_i \beta^k -
\frac{2}{3} \tilde{g}_{ij} \partial_k \beta^k, \nonumber\\ 
\lie_\beta \tilde{A}_{ij} & = &  \tilde{A}_{ij} \partial_k \tilde{A}_{ij} +
\tilde{A}_{ik} \partial_j \beta^k + \tilde{A}_{jk} \partial_i \beta^k -
\frac{2}{3} \tilde{A}_{ij} \partial_k \beta^k. \nonumber
\end{eqnarray*}

As in~\cite{Tichy:2006qn,Bruegmann:2006at}
we evolve the BSSN system as a partially constrained scheme,
where both the algebraic constraints  $\det(g) = 1$ and 
$\Tr(A_{ij})=0$ are enforced at every intermediate time step of the
evolution scheme. In addition, 
we also impose the first-order differential constraint
$\tilde{\Gamma}^i = - \partial_j \tilde{g}^{ij}$ by replacing
all undifferentiated occurrences of $\tilde{\Gamma}^i$ by
$- \partial_j \tilde{g}^{ij}$ instead of using the evolved
variable $\tilde{\Gamma}^i$.

Note that for puncture initial data the BSSN variable $\phi$
has a divergence of the form $\log r_A$
at each puncture. 
Since a logarithmic divergence is relatively weak, the moving
puncture approach consists of simply ignoring this divergence
by putting it between grid points at the initial time.
One option is to simply evolve the resulting initial data using
a finite differencing scheme, which effectively smooths
out any divergences, obviating the need for any
special treatment of the punctures. This is the approach
we have followed in our so called $P$-runs.
Another option is to replace the BSSN variable $\phi$
by a new variable~\cite{Campanelli:2005dd}
\begin{eqnarray}
\chi = e^{-4\phi} , \nonumber
\end{eqnarray}
which initially goes like $r_A^4$ at puncture $A$. We use this second option
in our $C$-runs.

The second ingredient in the moving-puncture method is a modification to the
gauge choice. We use a ``1+log'' lapse of the form~\cite{Campanelli:2005dd}
\begin{eqnarray}
	(\partial_t - \beta^i\partial_i) \alpha = -2\alpha K.\nonumber
\end{eqnarray}
For the shift, we use the gamma-freezing condition~\cite{Alcubierre02a,
Campanelli:2005dd}
\begin{equation}
\label{Gfreezing_ttt}
	\partial_t \beta^i = \frac{3}{4} B^i, \quad
	\partial_t B^i  =  \partial_t \tilde \Gamma^i - \eta B^i , 
\end{equation}
with $\eta=1.0/M$ for the $P$-runs. For the $C$-runs we used the modified
gamma-freezing condition
\begin{equation}
\label{Gfreezing_000}
(\partial_t - \beta^k\partial_k)\beta^i = \frac{3}{4} B^i, \quad
(\partial_t - \beta^k\partial_k) B^i  =  
(\partial_t - \beta^k\partial_k)\tilde \Gamma^i - \eta B^i ,
\end{equation}
where advection terms have been added to all time derivatives,
and where we choose $\eta=2.0/M$.


\subsection{Measurement of the mass and angular momentum}
\label{vol_int_method}

The black hole resulting from a BBH merger is defined by its mass and angular momentum, 
thus an accurate measurement of such global gauge-independent quantities becomes 
critical. One way to estimate these quantities is evaluating the ADM mass and angular
momentum after the merger. They are defined as surface 
integrals on a surface arbitrarily far from the system 
\cite{Murchadha:1974,Bowen:1980}
\begin{eqnarray}
\label{mass01}
M^{ADM} = {1 \over {16 \pi} } \oint_{\infty} (\partial_l g_{mr}-\partial_m g_{lr}) ~g^{nm} g^{lr} ~dS_n~,\\
J^{ADM}_i = {1 \over {8 \pi} } ~\epsilon_{il}^{~m} \oint_{\infty} x^l ~A^n_m ~dS_n~,
\label{ang01}
\end{eqnarray}
where $\epsilon_{ij}^{~k}$ is the Levi-Civita tensor and $dS_n \equiv {1 \over 2} \sqrt{g} ~\epsilon_{nlm} ~dx^l dx^m$.
The estimation of the final black hole parameters in numerical codes is currently 
done in several different ways: 1) by evaluating Eqs. (\ref{mass01}, \ref{ang01}) as
far as the grid size permits, 2) by measuring properties of the apparent,
event or isolated horizons (see for instance \cite{Schnetter:2006yt,
Campanelli:2006fy}), 3) by estimating the amount of emitted energy and
angular momentum in the form of gravitational radiation
or 4) by converting the surface integrals (\ref{mass01}, \ref{ang01}) to
volume integrals using Gauss' theorem. This last method, which has been
successfully employed in accretion disks around black holes (see for
instance \cite{Font:1998qr, Font:1998zq}), binary neutron star systems
\cite{Shibata:1999hn, Duez:2002bn, Marronetti:2003hx, Marronetti:2005bz,
Marronetti:2005aq} and single black hole spacetimes \cite{Yo:2002bm}, is
employed here and compared with results obtained from some of the other
techniques. Some of the advantages of this method are 1) the reduction of the 
influence of noise generated at the outer boundaries, 2) the reduction of gauge drift
effects, 3) it provides a real-time quality control factor at all times
during the simulation and 4) it complements, and sometimes improves, the
accuracy of alternative measurements. These advantages will be clarified in
the following section with examples and comparison with some of the
alternative methods. All the formulas used here are in Cartesian coordinates. 
The derivations in this section follow those of Yo
{\it et al.} \cite{Yo:2002bm} and Duez \cite{Duezprivate}.
 
Since we are interested in applying these equations in formalisms that perform a 
conformal decomposition of the physical metric $g_{ij} = e^{4 \phi} ~\bar{g}_{ij}$, 
it is useful to transform them accordingly. Following  \cite{Yo:2002bm}, Eqn. 
(\ref{mass01}) can be re-written as 
\begin{eqnarray}
M^{ADM} & = & {1 \over {16 \pi} }  \oint_{\infty}   
[e^{\phi}~(\partial_l \bar{g}_{mr}-\partial_m \bar{g}_{lr}) + 4 ~(\partial_l e^{\phi} ~ \bar{g}_{mr}-\partial_m e^{\phi} \bar{g}_{lr})] \nonumber\\
& &  ~~~~~~~~~~e^{\phi} ~\bar{g}^{nm} \bar{g}^{lr} ~d\bar{S}_n \nonumber\\
& = & {1 \over {16 \pi} }  \oint_{\infty} 
[\bar{g}^{lr} (\partial_l \bar{g}_{mr}-\partial_m \bar{g}_{lr}) -8 ~\partial_m e^{\phi}] ~\bar{g}^{nm} ~d\bar{S}_n \nonumber\\
& = & {1 \over {16 \pi} }  \oint_{\infty} (\bar{\Gamma}^n - \bar{\Gamma}^{ln}_{~~l} - 8~ \bar{D}^n e^{\phi}) ~d\bar{S}_n~,
\label{mass02}
\end{eqnarray}
with $d\bar{S}_n \equiv {1 \over 2} \sqrt{\bar{g}} ~\epsilon_{nlm} ~dx^l dx^m$, 
$\bar{\Gamma}^{i}_{jk}$  the conformal affine connections, 
$\bar{\Gamma}^i \equiv -\partial_j \bar{g}^{ij}$ and 
$\bar{D}_n$ the covariant derivative with respect to the conformal spatial metric. 
Similarly, Eqn. (\ref{ang01}) becomes
\begin{eqnarray}
\label{ang02}
J^{ADM}_i = {1 \over {8 \pi} } ~\epsilon_{il}^{~m} \oint_{\infty} x^l 
~\bar{A}^n_m ~d\bar{S}_n~.
\end{eqnarray}
Note that we have not assumed that $\sqrt{\bar{g}} = 1$, as is the case in the BSSN formalism.

The method studied in this paper converts straightforwardly these surface integrals 
using Gauss' law with the only provision of excluding the parts of the grid immediately 
surrounding the black holes. For an arbitrary vector field $f_i$, Gauss' law adopts the form
\begin{equation}
\label{gauss}
\oint_\infty f^n ~dS_n = \int_{V_\infty} \partial_n (\sqrt{g}~f^n) ~dx^3 + \sum_k \oint_{\partial \Omega_k} f^n ~dS_n~,
\end{equation}
where the first term represents the volume integral over all space except the parts 
enclosed by the closed surfaces $\partial \Omega_k$ that surround each one of the $k$ 
black holes. Numerical simulations like ours are performed using grids that cover a 
finite spatial volume $V$. Because of that, we can only provide estimates to the 
quantities defined in Eqs. (\ref{mass02}, \ref{ang02}). We will call the calculations 
of the mass and angular momentum performed in our finite grid volumes $M_V$ and $J_V$
respectively. After the merger, the gravitational fields settle down in the Kerr 
geometry corresponding to the final black hole. Once this occurs, the values of $M_V$ 
and $J_V$ should approximate the corresponding mass and angular momentum of the 
black hole remnant. The convergence of this approximation is studied in the next section.

In the case of the mass formula (\ref{mass02}), the direct application 
of (\ref{gauss}) plus some algebra leads to
\begin{eqnarray}
\label{mass03}
M_V & = & {1 \over {16 \pi} }  \int_{V}  
(\bar{R} + \bar{\Gamma}^n ~\bar{\Gamma}^{l}_{~nl} - \bar{\Gamma}^{lnm} \bar{\Gamma}_{nlm} - 
8~\bar{D}^2 e^{\phi}) ~\sqrt{\bar{g}} ~d^3x \nonumber\\
& & + {1 \over {16 \pi} }  \sum_k \oint_{\partial \Omega_k}
(\bar{\Gamma}^n - \bar{\Gamma}^{nl}_{~~l} - 8~ \bar{D}^n e^{\phi}) ~d\bar{S}_n~,
\end{eqnarray}
where $\bar{R}$ is the spatial Ricci scalar, $\bar{D}^2 \equiv \bar{g}^{mn} 
\bar{D}_m \bar{D}_n$, and the infinite volume $V_\infty$ has been replaced by the finite
volume $V$. The final step in the derivation of the mass formula is the 
use of the Hamiltonian constraint 
\begin{eqnarray}
\bar{D}^2 e^{\phi} = {e^\phi \over 8} \bar{R} + {e^{5\phi} \over 12} K^2 
- {e^{5\phi} \over 8} ~\bar{A}^{mn} \bar{A}_{mn}~, \nonumber
\end{eqnarray}
to eliminate in Eqn. ({\ref{mass03}) the term proportional to $\bar{D}^2 e^{\phi}$ 
\begin{eqnarray}
\label{mass04}
M_V & = & {1 \over {16 \pi} }  \int_{V}  
[(1-e^\phi)~\bar{R} + \bar{\Gamma}^n ~\bar{\Gamma}^{l}_{~nl} - \bar{\Gamma}^{lnm} 
\bar{\Gamma}_{nlm} + \nonumber\\
& & ~~~~~~~~e^{5 \phi}(~\bar{A}^{mn} \bar{A}_{mn} - {2 \over 3} K^2 ) ] 
~\sqrt{\bar{g}} ~d^3x \nonumber\\
& & + {1 \over {16 \pi} }  \sum_k \oint_{\partial \Omega_k}
(\bar{\Gamma}^n - \bar{\Gamma}^{nl}_{~~l} - 8~ \bar{D}^n e^{\phi}) ~d\bar{S}_n~.
\end{eqnarray}
In a similar manner, Eqn. (\ref{ang02}) is converted to
\begin{eqnarray}
\label{ang03}
J_{iV} & = & {1 \over {8 \pi} } ~\epsilon_{il}^{~m} \int_{V}
\left[ e^{6 \phi} \bar{A}^l_{~m} + x^l \bar{D}_n (e^{6 \phi} \bar{A}^n_{~m}) - {1 \over 2} e^{6 \phi} x^l \bar{A}_{ns} \partial_m \bar{g}^{ns}\right] ~d^3x + \nonumber\\
& & {1 \over {8 \pi} } ~\epsilon_{il}^{~m} \sum_k \oint_{\partial \Omega_k} x^l ~\bar{A}^n_m ~d\bar{S}_n~.
\end{eqnarray}
The momentum constraint 
\begin{eqnarray}
\bar{D}_n (e^{6 \phi} \bar{A}^n_{~m}) = {2 \over 3} ~e^{6 \phi} \bar{D}_m K~, \nonumber
\end{eqnarray}
is now used in Eqn. (\ref{ang03}), resulting in
\begin{eqnarray}J_{iV} & = & {1 \over {8 \pi} } ~\epsilon_{il}^{~m} \int_{V}
e^{6 \phi} \left[ \bar{A}^l_{~m} + {2 \over 3} ~x^l \bar{D}_m K - {1 \over 2} x^l \bar{A}_{ns} \partial_m \bar{g}^{ns}\right] ~d^3x + \nonumber\\
& & {1 \over {8 \pi} } ~\epsilon_{il}^{~m} \sum_k \oint_{\partial \Omega_k} x^l ~\bar{A}^n_m ~d\bar{S}_n~.
\label{ang04}
\end{eqnarray}

Our calculations were based on the BSSN formalism
for which the conformal transformation $g_{ij} = e^{4 \phi} ~\tilde{g}_{ij}$
is such that $\phi \equiv \ln(\tilde{g}^{1/12})$. This condition imposes the
algebraic constraints $\tilde{g} =1$ and $\tilde{\Gamma}^l_{~nl} = 0$ which
can be used to simplify even more Eqns. (\ref{mass04}) and
(\ref{ang04}). These equations are the formulas used in this article with the 
``bar" fields replaced by the corresponding BSSN (``tilde") counterparts.

Figure \ref{Bamgrid} shows the schematics of a typical grid structure. The
darker shading indicates higher resolution. In order to avoid the coordinate
singularity in the calculation of $M_V$ and $J_V$, we exclude the region
surrounding the black holes. Given BAM's mesh structure, it is particularly
simple to choose the outer edge of one of the moving boxes (the innermost in
Fig. \ref{Bamgrid}) as the boundary $\partial \Omega_k$. In the next section
we provide results for different choices of these boundaries. Both numerical
integrations (volume and surface) are performed using an extended version of
the trapezoidal rule with fourth order convergence \cite{NumRec}.

\begin{figure}
\centering
\vskip 1.0cm
\includegraphics[angle=-90,width=3in]{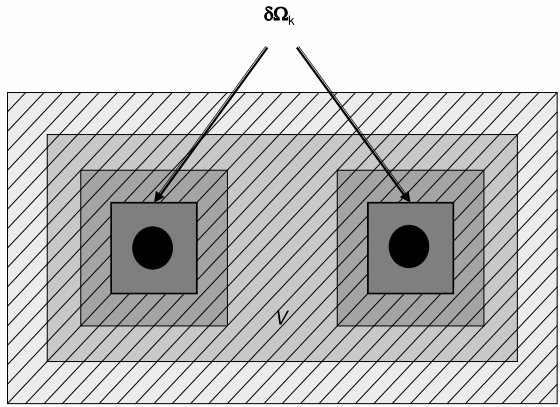}
\caption{Schematic diagram of BAM's grid structure.}
\label{Bamgrid}
\end{figure}


\section{Code performance}
\label{Perf}

The numerical results discussed in this paper were obtained with the BAM
code~\cite{Bruegmann:2003aw,Bruegmann:2006at}.
This code is based on a method of lines approach using 
fourth order finite differencing in space and 
explicit fourth order Runge-Kutta (RK) time stepping.
For efficiency, Berger-Oliger type mesh refinement is
used~\cite{Berger84}. 
The numerical domain is represented by a hierarchy of nested Cartesian
boxes. The hierarchy consists of $L+1$ levels of refinement,
indexed by $l = 0, \ldots, L$. A refinement level consists
of one or two Cartesian boxes with a constant grid-spacing 
$h_l = h_0/2^l$ on level $l$. We have used here $L=9$ to $11$ for the 
number of refinement levels, with the levels 0 through 5 each 
consisting of a single fixed box centered on the
origin (the center of mass). On each of the finer levels 6 through 
$L$, we initially use two sets of moving boxes centered on each 
black hole. When the black holes get close enough that
two of these boxes start touching, they are replaced by a single
box. The position of each hole is tracked by integrating 
the shift vector. We have used this same set up with 
different resolutions to perform convergence tests.
The notation used to describe these grid setups is as follows: 
the $C1$ run is represented by
\begin{itemize}
\item $C$1: $\chi_{\eta=2}[5\times 40:6\times 80][h_{10}=M/56.9:OB=729M]$
\end{itemize}
where $\chi$ represents the use of that dynamical variable, $\eta$ is the
free parameter in the shift vector formula,
$5\times 40$ indicates that we have 5 levels with moving
boxes of $40\times 40/2\times 40/2$ points and 6 levels with non-moving
boxes of $80\times 80/2\times 80/2$ points
(the divisions by 2 are due to using quadrant symmetry).
The resolution at the finest level ($l=10$ in this case) is $h_{10}=M/56.9$
and the outer boundary is placed at $\sim 729M$ from the origin. 

Finally, we note that BAM is MPI parallelized. When $N$ processors
are used, each box on each refinement level is divided into $N$
equally sized sub-boxes with added ghostzones. Each of these 
sub-boxes is then owned and evolved by one processor. The
ghostzones are synchronized in the usual way after each evolution step.
In this way, each processor owns exactly one sub-box 
of every mesh refinement box, which optimizes load balancing 
since then each processor works on the same number of grid points.
For additional details about the version of the BAM code used here,
see~\cite{Bruegmann:2006at}.

We tested the BAM code by running simulations of a high-spin black hole
binary system with individual spins $S/m^2 = 0.75$ aligned with the orbital
angular momentum. We choose $x-y$ as the binary's orbital plane which leaves
the $z$ component of the angular momentum as the only non-zero component of
$J_V$. Table \ref{table_runs} shows the characteristics of the simulations
performed for this article. The runs are grouped in those using $\chi$ 
($C$-runs) as the dynamical variable and those using $\phi$ ($P$-runs). For 
the former (latter), we chose a value for the shift parameter $\eta$ of $2.0/M$ 
($1.0/M$). In general, the simulations performed here have higher maximum 
spatial resolutions than those in \cite{Bruegmann:2006at}, showing the moving 
punctures method's stability even when the coordinate separation between the 
grid points and the punctures is as small as $M/320$ (run $P6$) \footnote{See 
\cite{Hannam:2006vv, Baumgarte:2007ht} for details on the spacetime geometry close 
to the punctures.}. Another difference is that the grids used for the $C$-runs are 
larger than those of  \cite{Bruegmann:2006at}. These extensions of the grid size 
and high resolutions did not excite any undesirable instabilities; in the case of
$C2$, the simulation was run for the equivalent of over 20 orbital periods
\footnote{Assuming a nominal orbital length of 114M, obtained from the initial
data set angular velocity.}.

\begin{table}
\centering
\begin{tabular}{c|c}
$Run$	& $Grid$  \\
\hline
\hline
C1  & $\chi_{\eta=2}[5\times 40:6\times 80]~~[h_{10}=M/56.9 :OB=729M]$\\
C2  & $\chi_{\eta=2}[5\times 44:6\times 88]~~[h_{10}=M/62.6 :OB=728M]$\\
C3  & $\chi_{\eta=2}[5\times 48:6\times 96]~~[h_{10}=M/68.3 :OB=727M]$\\
C4  & $\chi_{\eta=2}[5\times 52:6\times 104]~[h_{10}=M/73.9 :OB=727M]$\\
C5  & $\chi_{\eta=2}[5\times 56:6\times 112]~[h_{10}=M/79.6 :OB=726M]$ \\
P1  & $\phi_{\eta=1}[4\times 32:6\times 64]~~[h_9=M/68.3 ~:OB=244M]$ \\
P2  & $\phi_{\eta=1}[4\times 40:6\times 80]~~[h_9=M/85.3 ~:OB=243M]$ \\
P3  & $\phi_{\eta=1}[4\times 48:6\times 96]~~[h_9=M/102.4 :OB=243M]$ \\
P4  & $\phi_{\eta=1}[4\times 56:6\times 112]~[h_9=M/119.5 :OB=242M]$ \\
P5  & $\phi_{\eta=1}[4\times 32:8\times 64]~~[h_{11}=M/68.3 :OB=975M]$ \\
P6  & $\phi_{\eta=1}[4\times 40:6\times 80]~~[h_{9}=M/160.0 :OB=130M]$\\
\end{tabular}
\caption{\label{table_runs}
Grid structure of the $\chi$ ($C$) and $\phi$ ($P$) and runs. The $\chi$ 
($\phi$) runs used $\eta=2.0/M$ ($\eta=1.0/M$).}
\end{table}

One of the strongest characteristics of the BAM code is its ability to
perform good quality simulations with modest computer resources. Table
\ref{table_perf} shows typical running times and memory requirements for the
simulations of Table \ref{table_runs}. Note that BAM performs
faster after the merger, when only one set of non-moving boxes remains. All
the simulations presented in this paper were run on dual processor
workstations, namely a AMD Dual Opteron 2.2GHz workstation with 8Gb of memory
and a Intel Dual Xeon 2.6GHz workstation with 16Gb. The former computer can be 
purchased at the time of publication for less than \$3000 USD. Note that, 
while none of the runs presented here required significantly more than 8Gb 
of memory, the cost of expanding the workstation memory capabilities has 
dropped considerably in the past year. Currently, a 1Gb memory module for 
our AMD machine retails for about \$100 USD.

Evolutions of BBH that start at larger separations would in principle 
require longer running times and more computer memory. Estimates of how 
much longer it would take to run extra orbits on any of our example runs 
can be obtained from the information given in the last column of Table 
\ref{table_perf}. The memory requirements, on the other hand, will depend 
on the grid structure to be used and obviously the location of the outer 
boundaries. The memory requirements for runs such as ours would not change 
as long as the separation distance is not increased to more than $9.5M$, 
which would yield an initial orbital period of more than $200M$, 
thus leading to several more orbits before merger. The runs described in 
Table \ref{table_runs} were performed using quadrant symmetry. However, generic 
BBH with arbitrary masses and spins have to be simulated in full grids
which, all things kept equal, would require about four times more memory 
and execution time. These requirements, however, can be reduced by adopting 
a different grid geometry. In Tichy and Marronetti \cite{Tichy:2007hk}, 
generic BBH runs were performed using a grid 
$\chi_{\eta=2}[5 \times 48:5 \times 54]~~[h_9=M/56.9 :OB=240M]$. The 
performance details of one of these runs (completed on the Dual Intel 
Workstation) have been added in the last row of Table \ref{table_perf} where
we see that, while slower than the previous runs, this simulation only
takes a couple of days of execution time per orbit. 

Finally we performed the following comparisons. Firstly, we performed 
run $C5$ also on a supercomputer (Cray XT3 MPP system at the Pittsburgh 
Supercomputer Center) using 32 processors. The execution on this machine 
was about 7 times faster ($20~M/hr$) than on our AMD Dual Opteron 
workstation. Secondly, we have also evolved model $C5$ of Table 
\ref{table_runs} with the LEAN code \cite{Sperhake:2006cy}. Because 
quadrant symmetry is not implemented in the current version of the grid 
driver of the LEAN code, this simulation was performed using 
equatorial symmetry, but using four processors. The results are listed as 
$LC5$ in Table \ref{table_runs}. We emphasize a few 
caveats in this last comparison. For instance, it was not possible to use 
identical grid setups due to the different types of symmetries used 
and the use of cell-centered and vertex-centered grids in BAM and LEAN 
respectively. Furthermore, the codes continue to undergo further 
development with likely improvements, in particular in the case of memory 
usage in BAM. Finally, the results are likely to be affected by the 
inclusion of further diagnostic tools, such as horizon finding. However, 
BAM and LEAN's performance are similar, both in terms of memory usage 
and speed, indicating that the critical aspect of these codes efficiency 
is the particular grid structure, more than in intrinsic coding details 
of the evolution equations.

\begin{table}
\centering
\begin{tabular}{c|c|c|c|c|c}
$Run$	 & $Mem$ & $Pre-merger$ & $Post-merger$ & $One ~orbit$ \\
\hline
\hline
C1  & $4.2~Gb$ & $10.8~M/h$ & $17.8~M/h$ & $0.44~d$\\
C2  & $5.5~Gb$ & $7.7~M/h$ & $12.9~M/h$ & $0.62~d$\\
C3  & $6.3~Gb$ & $5.2~M/h$ & $9.4~M/h$ & $0.91~d$\\
C4  & $7.7~Gb$ & $3.6~M/h$ & $6.6~M/h$ & $1.32~d$\\
C5  & $8.1~Gb$ & $3.1~M/h$ & $5.8~M/h$ & $1.53~d$\\
P1  & $4.1~Gb$ & $7.2~M/h$ & $10.1~M/h$ & $0.66~d$\\
P2  & $4.3~Gb$ & $6.8~M/h$ & $9.8~M/h$ & $0.70~d$\\
P3  & $4.9~Gb$ & $6.2~M/h$ & $9.1~M/h$ & $0.78~d$\\
P4  & $8.2~Gb$ & $1.7~M/h$ & $3.1~M/h$ & $2.79~d$\\
P5  & $4.6~Gb$ & $6.4~M/h$ & $9.0~M/h$ & $0.74~d$\\
LC5 & $6.5~Gb$ & $2.3~M/h$ & $4.6~M/h$ & $2.07~d$\\
FG  & $11.7~Gb$ & $2.3~M/h$ & $4.6~M/h$ & $2.20~d$\\
\end{tabular}
\caption{\label{table_perf}
Typical performance of the BAM code on a AMD Dual Opteron 2.2GHz workstation
(runs $P4$, $C1$, $C3$ and $C5$,) and an Intel Dual Xeon 2.6GHz workstation 
(runs $P1$, $P2$, $P3$, $P5$ and $C4$).
While these workstations are by themselves not particularly expensive, they
were fitted with $8Gb$ (Opteron) and $16Gb$ (Xeon) of memory 
that increased their cost significantly. The last column shows the time (in days) 
it takes to evolve for one orbit using a nominal orbital length of 114M, 
obtained from the initial data set angular velocity. Due to the loss of the files, 
$P6$ performance could not be estimated. $LC5$ corresponds to a simulation 
identical to $C5$ but performed using the LEAN code \cite{Sperhake:2006cy}. 
The last row ($FG$) corresponds to a simulation using full grid presented
in \cite{Tichy:2007hk}.}
\end{table}


\section{Code tests}
\label{Tests}

\subsection{Convergence tests}

Table \ref{table_results} indicates the values obtained for the mass and
angular momentum at the time the simulations were stopped ($t_F$). We also 
present the time of merger, estimated as the moment when the lapse function 
drops below a threshold value of 0.3. The error bars simply present the change 
of each quantity in the last 100M of the simulation (or a nominal value of
0.001, if the change is smaller than this threshold). The $C$-runs show
larger changes in $J_V$ than the $P$-runs due to the larger grid size
used in the former simulations which requires longer evolution times for the
volume integrals to settle.

\begin{table}
\centering
\begin{tabular}{c|c|c|c|r|r}
$Run$	 & $M_V/M$ & $J_V/M^2$ & $J_V/{M_V}^2$ & $t_F/M$ & $t_M/M$\\
\hline
\hline
C1  & $0.914\pm0.002$ & $0.575\pm0.005$ & $0.688$ & $1582$ & $262$\\
C2  & $0.914\pm0.001$ & $0.567\pm0.003$ & $0.679$ & $2371$ & $262$\\
C3  & $0.913\pm0.001$ & $0.623\pm0.007$ & $0.749$ & $1511$ & $262$\\
C4  & $0.912\pm0.001$ & $0.640\pm0.007$ & $0.784$ & $1518$ & $263$\\
C5  & $0.912\pm0.001$ & $0.653\pm0.007$ & $0.787$ & $1594$ & $263$\\
P1  & $0.914\pm0.001$ & $0.625\pm0.002$ & $0.748$ & $1368$ & $297$\\
P2  & $0.915\pm0.001$ & $0.636\pm0.001$ & $0.760$ & $1911$ & $291$\\
P3  & $0.913\pm0.001$ & $0.671\pm0.003$ & $0.805$ &  $869$ & $279$\\
P4  & $0.911\pm0.001$ & $0.682\pm0.005$ & $0.822$ &  $862$ & $276$\\
P5  & $0.917\pm0.001$ & $0.627\pm0.004$ & $0.746$ & $1187$ & $292$\\
P6  & $0.915\pm0.001$ & $0.631\pm0.006$ & $0.754$ &  $928$ & $-$\\
\end{tabular}
\caption{\label{table_results}
Values of the mass $M_V$ and angular momentum $J_V$ at the time the simulations
from Table \ref{table_runs} were stopped ($t_F$). The merger time ($t_M$) is
estimated from the time when the minimum value of the lapse drops below 0.3.
The error bars are estimated from the variation in the last $100M$.
Due to the loss of the files, $P6$ merger time could not be estimated.}
\end{table}

Figure \ref{InnSurf} shows a comparison of $M_V$ and $J_V$ for different
positions of the inner surface for run $C5$. The length of the inner cube
side ($d$) is varied in multiples of $d=0.79M$. Note for reference that the
coordinate radius of the final black hole is $\sim 0.73M$.
The curves obtained for the smallest cube sizes show noise during
the pre-merger stages which disappears when increasing the cube side. The
spikes present at this stage are a numerical artifact caused by the crossing
of the moving boxes of the $x-z$ symmetry plane. After the merger, the curves
settle to values that agree within a relative error of less that $0.2\%$ for
$M_V$ and $0.04\%$ for $J_V$. This seems to indicate that, when measuring the
characteristics of the final black hole, the position of the inner cube does
not affect greatly the results.

\begin{figure}
\centering
\vskip 1.0cm
\includegraphics[angle=0,width=3.5in]{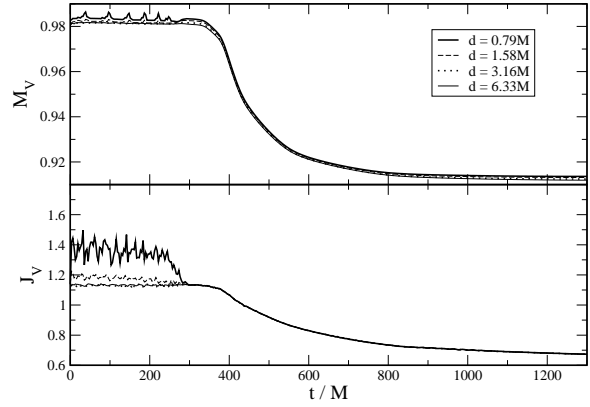}
\caption{
Comparison of $M_V$ and $J_V$ for different inner surfaces.
The curves correspond to
run $C5$. The side of the inner surface cube is denoted by $d$.}
\label{InnSurf}
\end{figure}

Figure \ref{BBox} compares the same quantities for two runs ($P1$ and $P5$)
that only differ in the size of the numerical grid. By adding two extra
refinement levels to the outside of the grid used for $P1$, we moved the
outer boundaries out by a factor of four. The main difference between these
$P$-runs is the way the curves relax to their final values which agree to
within a relative error of $0.01\%$ ($0.1\%$) for $M_V$ ($J_V$). Again, this
seems to show that the size of the grid does not affect significantly
the final values of the global quantities.

\begin{figure}
\centering
\vskip 1.0cm
\includegraphics[angle=0,width=3.5in]{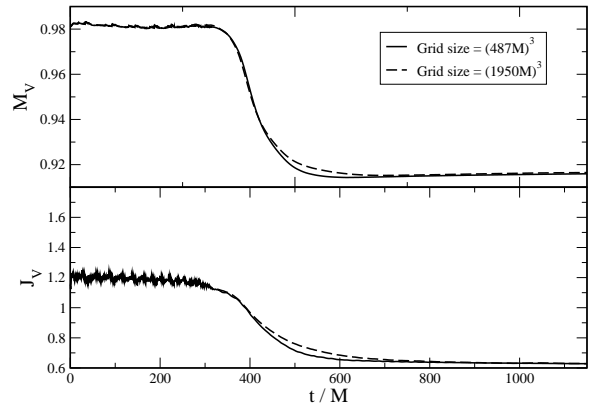}
\caption{
Comparison of $M_V$ and $J_V$ for two runs with different
grid sizes. The curves
correspond to runs $P1$ (solid) and $P5$ (dashed). $P5$ is identical to
$P1$, except for the presence of two additional low resolution outer levels
which extend the outer boundaries by a factor of four.}
\label{BBox}
\end{figure}

Figure \ref{Resol} shows a comparison of runs where the maximum grid
resolution has been varied, leaving the rest of the grid characteristics
identical. Note that the $J_V$ curves present a downward trend that persists
well after the merger. This trend gets less pronounced for higher resolution
and might be related to a similar coordinate drift observed in
\cite{Bruegmann:2006at}. Fig. \ref{Resol_Conv} presents convergence plots
for the runs $C2$ to $C5$. The runs are grouped in two sets {$C2$, $C4$, and
$C5$} (thick upper curves) and {$C3$, $C4$, and $C5$} (thin lower curves) 
and their differences are compared and scaled according to a putative fourth 
order convergence. The expected fourth order convergence is approached better 
by the set with higher resolution.

\begin{figure}
\centering
\vskip 1.0cm
\includegraphics[angle=0,width=3.5in]{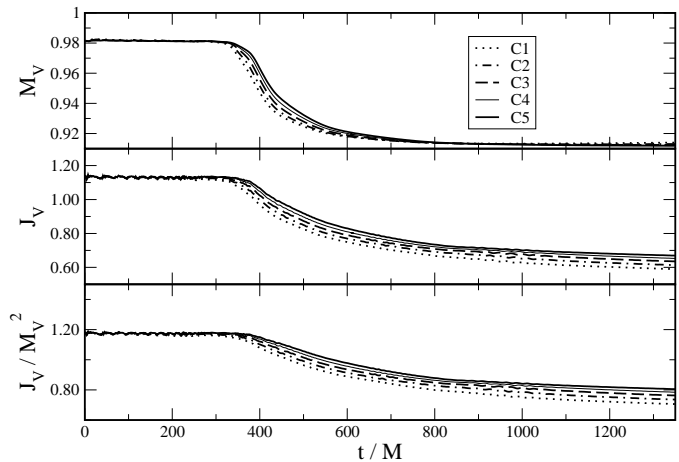}
\caption{
Comparison of $M_V$ and $J_V$ for runs with different grid resolutions. 
The curves correspond to runs $C1$ to $C5$. 
The values of $M_V$ and $J_V$ used for this test
correspond to inner surface cubes of size $6.33M$.}
\label{Resol}
\end{figure}

\begin{figure}
\centering
\vskip 1.0cm
\includegraphics[angle=0,width=3.5in]{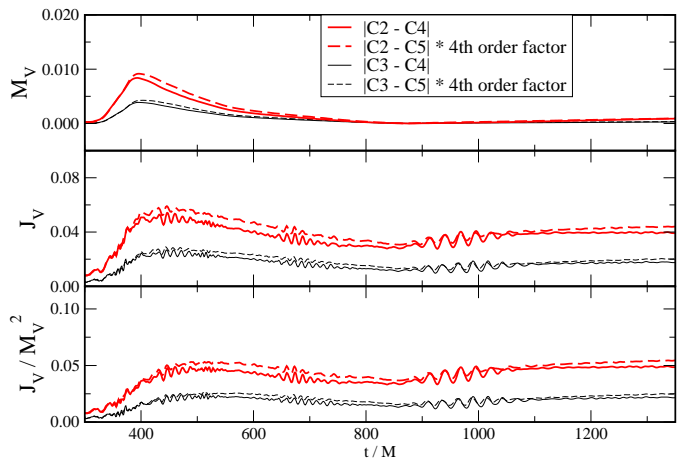}
\caption{
Convergence of $M_V$ and $J_V$ with grid resolution.
The curves correspond to runs $C2$ to $C5$. 
The values of $M_V$ and $J_V$
used for these plots correspond to inner surface cubes of size $6.33M$.}
\label{Resol_Conv}
\end{figure}

We also compare two runs with different characteristics but identical
maximum grid resolution. The curves from Fig. \ref{PvsC} correspond to runs
$P1$ (solid) and $C3$ (dashed). $P1$ was performed using $\phi$ as the
dynamical variable, Eq.~(\ref{Gfreezing_ttt}) with $\eta=1.0/M$ in the recipe
for the shift and it had outer boundaries placed at $244M$. $C3$ used $\chi$
as the dynamical variable, Eq.~(\ref{Gfreezing_000}) with $\eta=2.0/M$ and outer
boundaries at $727M$. The values of $M_V$ and $J_V$ used for these plots
correspond to inner surface cubes of size $6.33M$. As in Fig. \ref{BBox},
the difference in relaxation is mostly due to the difference in grid sizes.
The relative difference between the global quantities at
the end of these runs is $0.1\%$ for $M_V$ and $1\%$ for $J_V$.

\begin{figure}
\centering
\vskip 1.0cm
\includegraphics[angle=0,width=3.5in]{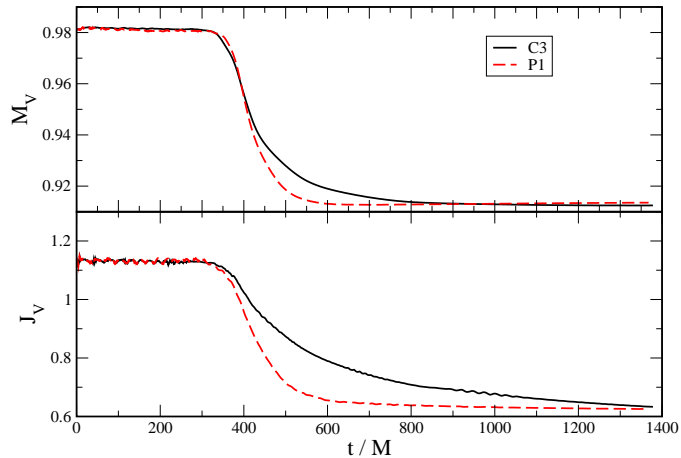}
\caption{
Comparison of $M_V$ and $J_V$ from two runs using the dynamical 
variables $\phi$ ($P1$) and $\chi$ ($C3$).}
\label{PvsC}
\end{figure}

Finally, in order to determine the values of the mass and angular momentum 
for the $C$-runs of Table \ref{table_runs}, we used Richardson extrapolation 
with a polynomial of the type
\begin{eqnarray}
\label{rich}
P(h_{0}) = A_0+A_1*h_{0}^4+A_2*h_{0}^5~.
\end{eqnarray}
Note that this formula assumes fourth order convergence which, according to
Fig. \ref{Resol_Conv}, is only approximate for the runs of this
paper. The extrapolation is performed at $t=1500M$ from runs $C3$, $C4$, and
$C5$, giving $M_V=0.909$ and $J_V=0.753$. These values agree to 
within $3\%$ of those reported in \cite{Campanelli:2006uy}. 


\subsection{Comparison of $M_V$ and $J_V$ with alternative estimates of 
the mass and angular momentum of the final black hole}
\label{Comp}

Alternative estimations of the mass and angular momentum of the final black 
hole can be derived from evaluating the surface integrals (\ref{mass01}, 
\ref{ang02}) at finite radii. Figure \ref{VvsS} compares the values of 
$M_V$ and $J_V$ with integrations performed on spherical surfaces at 
coordinate radii $30M,~50M$ and $100M$. The mass curves corresponding to 
the spherical surface integrations have very large errors that behave in a 
non-systematic way when the extraction radius is increased.
At the same time, the calculations of the angular momentum using 
spherical surface integrals, while noisy before the merger, agree with $J_V$ 
to within a $0.001\%$ relative error at the end of the simulation.

\begin{figure}
\centering
\vskip 1.0cm
\includegraphics[angle=0,width=3.5in]{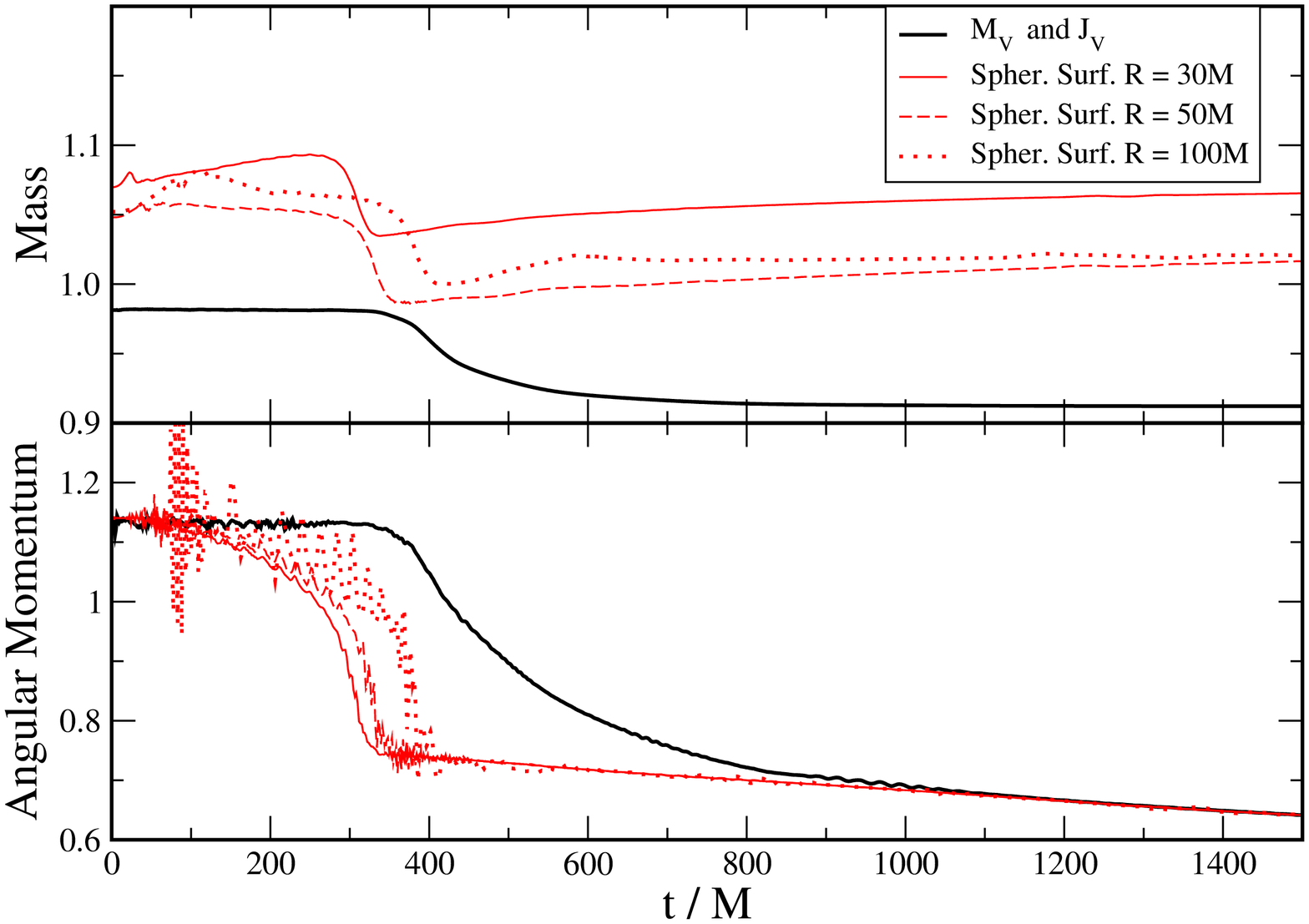}
\caption{
Comparison of $M_V$ and $J_V$ (solid) and spherical
surface integrations at radii $30M$ (dotted), $50M$ (dashed), and $100M$
(dashed-dotted). The curves correspond to run $C4$.}
\label{VvsS}
\end{figure}

Figure \ref{MvsE} shows the radiated energy calculated from the flux through
spherical surfaces using the Newman-Penrose scalar $\Psi_4$ (see Eq. (52) in 
\cite{Bruegmann:2006at}) at radii $15M$, $30M$, $60M$, $80M$ for run $C4$. From the
radiated energy, we generated an estimate for the time evolution
of the mass by subtracting those curves from the initial $M^{ADM}$ as
reported in Table \ref{punc_par_tab} (dashed curves).  These curves show a 
downward drift that is more pronounced with the proximity of the spherical 
surface to the center of the grid. The result obtained for radius $80M$ is the 
one that appears to be the least affected by this effect and it agrees with 
$M_V$ within a relative error of about $1\%$.

\begin{figure}
\centering
\vskip 1.0cm
\includegraphics[angle=0,width=3.5in]{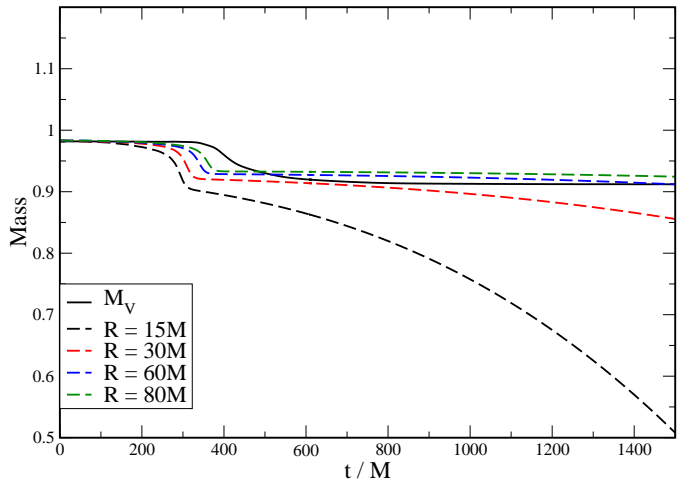}
\caption{
Comparison of $M_V$ (black solid) and mass
estimates from the energy radiated through spherical surfaces at
different radii (color dashed). The latter curves were generated by 
subtracting the radiated energy from the initial $M^{ADM}$.
The curves correspond to run $C4$ and inner surface cube size $6.33M$.}
\label{MvsE}
\end{figure}

The Christodoulou formula, valid for stationary axisymmetric spacetimes
like the Kerr geometry of the final black hole, gives the following relation
between the angular momentum $J$, mass $M$ and irreducible mass of a black hole
$M_{irr}$
\begin{eqnarray}
\label{christ}
J = 2 M_{irr} ~ \sqrt{{M}^2 - M_{irr}^2}~.
\end{eqnarray}
Here 
\begin{eqnarray}
M_{irr} = \sqrt{\frac{A_H}{16\pi}} \nonumber
\end{eqnarray}
is determined from the proper area $A_H$ of the apparent horizon (for the run
$C4$ $M_{irr} \approx 0.825M$). We verified relation (\ref{christ})
for several of the runs of Table \ref{table_runs} and found it to be satisfied 
to a relative error of less than $0.2\%$.


\section{Results and conclusions}
\label{Res}

The main goal of this paper is to show the ability of the BAM code to perform 
certain simulations of binary black holes with relatively modest computer 
resources. The simulations presented here were performed on
dual processor workstations that have been outfitted with at least 8Gb of
memory. Machines like these retail at the moment of publication for less than
\$3000 USD, making them easily accessible to any research group.

Our runs were based on the same initial state: one corresponding to two identical 
black holes with intrinsic spin parameters $S/m^2= 0.75$ and spins parallel to 
the orbital angular momentum and which started out at a coordinate separation of 
$6M$. This data set is similar to the one used by Campanelli \et 
\cite{Campanelli:2006uy}. Our results for the mass and angular momentum of the 
final black hole agree to within $3\%$ of those of \cite{Campanelli:2006uy}. 
Better accuracy could be 
achieved with higher resolution runs, however they would also demand larger
computer resources. We are currently testing different grid structures (i.e.,
varying moving boxes sizes, number, etc.) that improve
these runs accuracy without increasing the computational burden (see, for instance 
\cite{Tichy:2007hk}).
We performed high spatial resolution simulations (of up to $M/160$) and 
large grid size (up to outer boundaries at $975M$) to test the moving punctures 
method's stability and robustness. None of the runs showed any signs of
exponentially growing instabilities; they were stopped due to the long
real-time duration of the simulations. In one of our test cases, the simulation
was run for the equivalent of more than 20 orbital periods. 

The simulations discussed in this paper were performed using quadrant symmetry 
which require about four times less memory and computer time than non-symmetric
scenarios. An improved choice of grid structure can (to some extent) minimize
these requirements (see ~\cite{Tichy:2007hk}, for non-symmetric 
simulations using workstations). BAM capabilities
for efficient use of computer resources permit the exploration of BBH parameter
space by enabling low resolution simulations that only require workstations
or one or two nodes per run in local Bewoulf clusters, where many of such
runs can be done simultaneously. Ongoing code optimization is currently enhancing 
the code performance with regard to memory and cpu usage. 
Nevertheless, it is clear that today's most demanding binary simulations still 
require supercomputer resources. However, given the rapid growth of computer 
power and high efficiency codes such as BAM, even these simulations might be 
within the reach of workstation resources in the next couple of years.


\ack
Special thanks to B. Israel for help with running the BAM code. 
This work was partially supported by NSF Grant PHY0555644, 
National Computational Science Alliance under Grants PHY020007N. 
PHY050010T, PHY050015N and PHY060021P and by DFG grant SFB/Transregio~7 
``Gravitational Wave Astronomy''. We thank the DEISA Consortium 
(co-funded by the EU, FP6 project 508830), for support within the 
DEISA Extreme Computing Initiative (www.deisa.org);
Code development was also performed at LRZ Munich and HLRS, Stuttgart.
J.G. and U.S. acknowledge support from the ILIAS Sixth Framework 
Programme.


\end{document}